# Planetary Science Virtual Observatory architecture

**S. Erard [1], B. Cecconi [1], P. Le Sidaner [2], J. Berthier [3], F. Henry [1], C. Chauvin[2], N. André[4], V. Génot[4], C. Jacquey[4], M. Gangloff[4], N. Bourrel[4], B. Schmitt[5], M. T. Capria[6], G. Chanteur[7]**

***[1] LESIA, Observatoire de Paris/CNRS/UPMC/Univ. Paris-Diderot***
*5 pl. J. Janssen 92195 Meudon, France. email: stephane.erard@obspm.fr*

***[2] DIO-VO, UMS2201 CNRS, Observatoire de Paris***
*61 av. de l'Observatoire, 75014 Paris, France*

***[3] IMCCE, Observatoire de Paris/CNRS***
*61 av. de l'Observatoire, 75014 Paris, France*

***[4] CDPP, IRAP/CNRS/Univ. Paul Sabatier***
*9 avenue du colonel Roche, 31068 Toulouse, France*

***[5] INAF/IAPS, Rome, Italy***

***[6] LPP / CNRS, Ecole Polytechnique***
*route de Saclay, 91128 Palaiseau cedex, France*

***Corresponding author:*** *S. Erard, LESIA, Observatoire de Paris*
*5 pl. J. Janssen 92195 Meudon, France.*
*email: stephane.erard@obspm.fr*
*Tel : (33) 1 45 07 78 19*

## ABSTRACT

In the framework of the Europlanet-RI program, a prototype of Virtual Observatory dedicated to Planetary Science was defined. Most of the activity was dedicated to the elaboration of standards to retrieve and visualize data in this field, and to provide light procedures to teams who wish to contribute with on-line data services. The architecture of this VO system and selected solutions are presented here, together with existing demonstrators.

## INTRODUCTION

After more than 10 years of development, the astronomical Virtual Observatory (VO) has reached a point of maturity, and some of its tools are now routinely used in research activities. For historical and organizational reasons, Planetary Science was only marginally involved in this process. However, in the recent years several Planetary Science oriented projects were started, with various scopes and objectives: PDS (Planetary Data System, the NASA Planetary Science data archive) has developed several data access systems, and the PDS4 system is explicitly oriented towards easy user access; the IPDA (International Planetary Data Alliance, gathering most space agencies) has started the definition of a VO system to inter-connect space-borne data archives more generally. The planetary plasmas community also organized itself around specific data handling tools, such as AMDA at CDPP, Toulouse. In parallel, several programs funded by the European Union focused on data services and data processing related to Solar System studies: HELIO, IMPEx, VAMDC, CASSIS, and Europlanet-RI.

Among many activities, the Europlanet-RI program included a data access activity called IDIS. One of its topics was to study the Planetary Science VO and to produce a demonstrator of such a data system. The architecture of this project was defined in a Joint Research Activity (JRA) and the first data services were implemented in a Service Activity (SA), globally referred to as IDIS (Integrated and Distributed Information Services). The main actors in this process were VO-Paris Data Centre (a structure of the Observatory of Paris involved in Virtual Observatory activities [1]), CDPP (the planetary plasma data center in Toulouse), IAPS (an INAF laboratory in Rome), IPAG (a laboratory in Grenoble involved in spectroscopy in solid phase, also partner of the VAMDC program), and IWF in Graz. The overall infrastructure defined in Europlanet-RI is described here, with a stress on the solutions adopted and possible developments to be implemented in the future Europlanet_2020-RI program (currently under definition).

The aim of the activity was to facilitate searches of Solar System related data in big archives and sparse databases, to make on-line data access and visualization possible, and to allow small data providers to share their data in an interoperable environment with minimum effort. This system makes intensive use of previous studies and developments led in Astronomy (IVOA) and by space-borne data archive services (IPDA, PDS), as well as in solar Physics (HELIO program) and spectroscopy (VAMDC program). In particular, it remains consistent with extensions of IVOA standards.

The global architecture involves existing data services accessible through IVOA protocols (Cone Search, TAP…) or the IPDA protocol (PDAP) whenever relevant. However, a more general standard has been devised to handle the specific complexity of Planetary Science, e.g. in terms of measurement types and coordinate frames. This protocol, named EPN-TAP, is based on Table Access Protocol (TAP) and includes precise requirements to describe the contents of a data service [see companion paper: Erard et al. this issue]. The data services are declared in the extended IVOA registry based at VO-Paris, which is queried by a web-based client handling both EPN-TAP and PDAP protocols. Since the original funding was European, some priority targets are related to European research activities – however, the system is not proprietary and is intended to fulfill the needs of the community in a broad sense.

The client itself and some demonstrators are available from the VO-Paris Europlanet node at http://voparis-europlanet.obspm.fr/ (acronyms and web links are provided in annex).

## PROPOSED ARCHITECTURE

A general scheme is described in Fig. 1, which illustrates the sequence of steps in a typical working session. The user is working at his computer, sends queries to databases to identify data of interest, and gets answers. The data can then be loaded in memory, plotted in various forms (images, spectra…), and are possibly sent to more elaborated tools performing specialized functions or processing. These various steps are commented below.

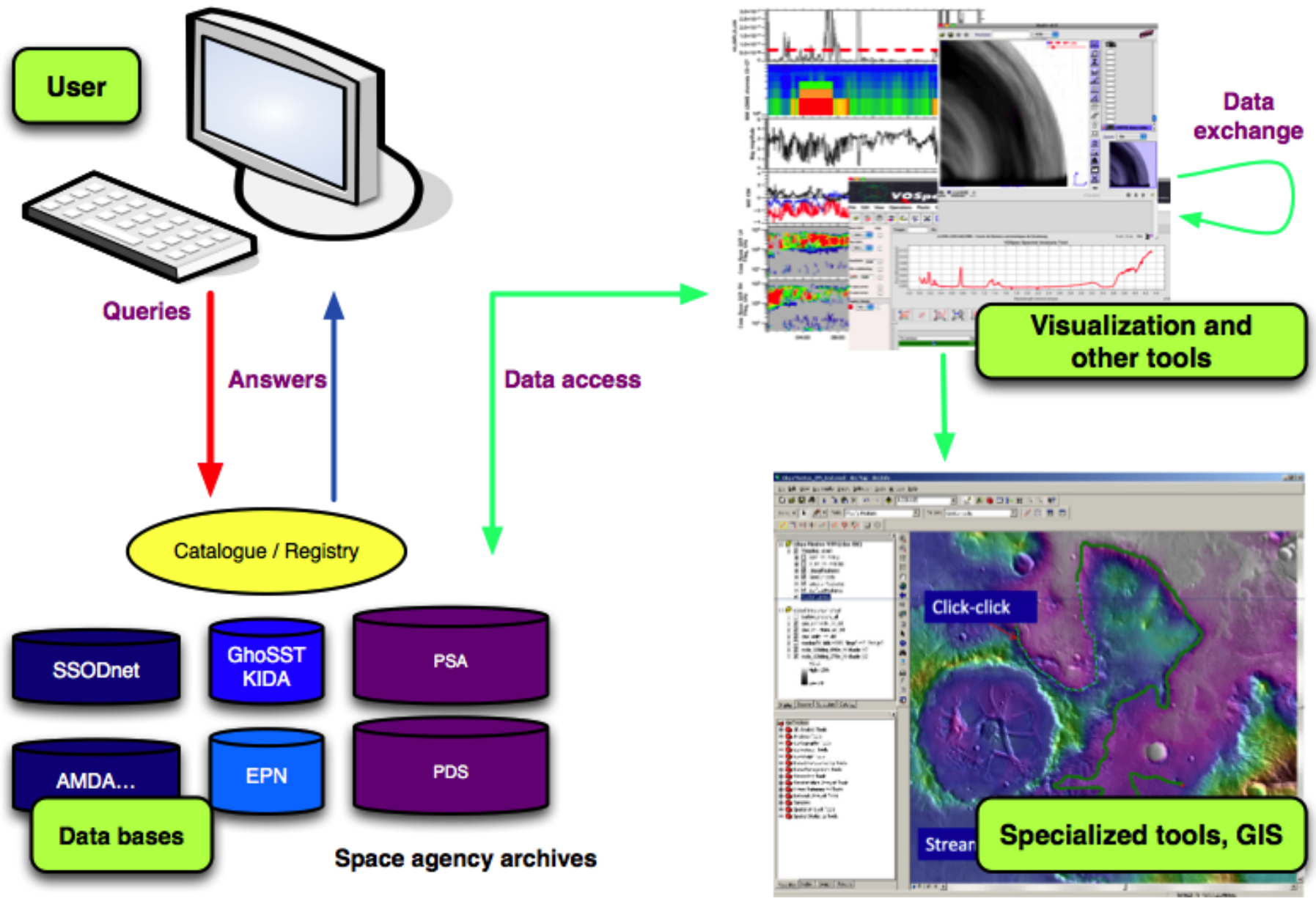

Figure 1: Overall scheme of the Planetary Science VO

## DATA SCOPE

The perimeter of data to be accessed by the Planetary Science VO derives from the objectives stated in the Europlanet program proposal. It includes (Fig. 2):

- Data bases produced by various work packages during the Europlanet program (JRA4/task4).
- A selection of space-borne data from planetary missions. This includes data from European space missions, i.e. access to ESA's Planetary Science Archive (PSA) [2], but also calibrated or reduced data sometimes available only in instrument teams.
- Specialized data services and tools related to participants of the Europlanet / IDIS activity are also linked to the system, e.g. GhoSST (Grenoble Astrophysics and Planetology Solid Spectroscopy Thermodynamics) at IPAG, AMDA (Automated Multi Dataset Analysis) at CDPP/Toulouse or SSODnet (Solar System Open Database Network) at VO-Paris. This also includes services designed in the frame of related European programs, such as HELIO and IMPEx.
- Big data repositories that include Planetary Science data and predate Europlanet are other natural targets to expand this system, e.g. the ESO and HST archives.
- Data sets directly published in a compliant form by data providers, typically as end product of a research activity, after scientific publication.

The Planetary Science VO is required to be open so as to allow external data providers to include their databases in the system with minimum efforts. This includes observational data derived from space missions or ground-based telescopes, but also reference data acquired in the laboratory, and simulations.

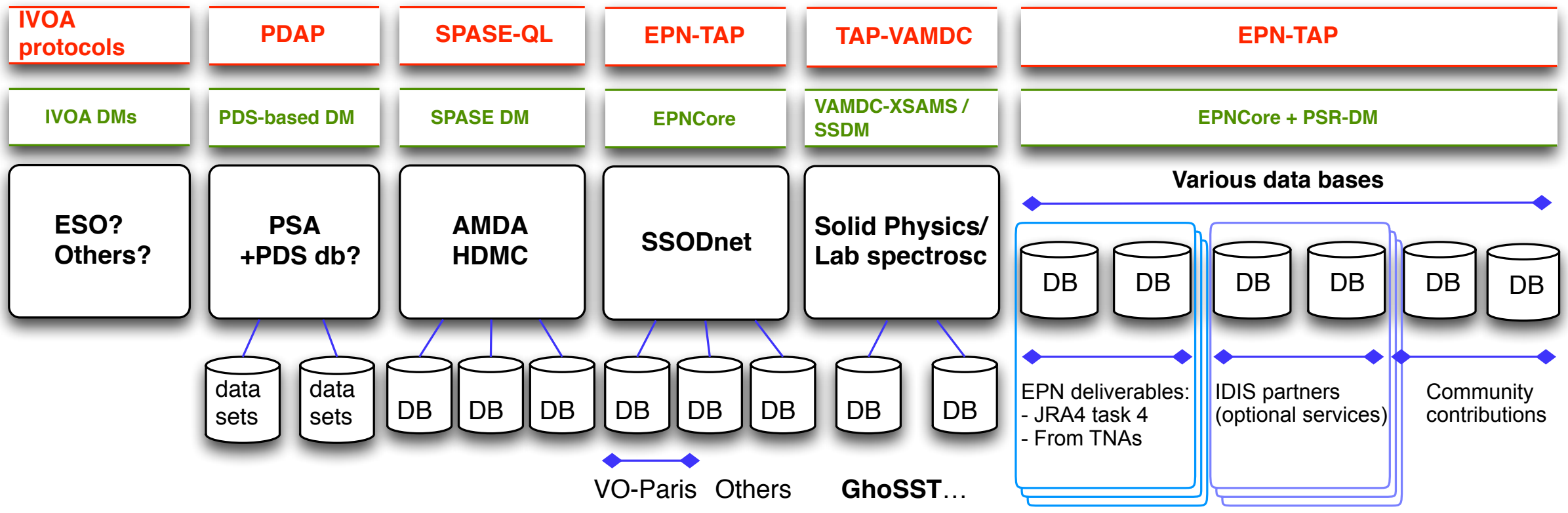

Figure 2: Data services expected to be included in the Europlanet VO, with related access protocols (red) and data models (green)

The VO system is therefore intended to provide access to a variety of data related to planetary surfaces, atmospheres, and interiors, laboratory samples (including meteorites), Solar System plasmas, and possibly solar data… It is stressed that there is a high level of heterogeneity in these data, in particular with respect to astronomical data:

- Formats: although most data formats are readily handled, PDS3-formatted data from space missions are difficult to read due to the lack of standard/versatile software, and require a special input process. Many specific formats are also in use, e.g. CDF in plasma physics.
- Coordinate frames: sky images exist, but in contrast with astronomy the targets are moving and can be identified from their coordinates only at a given time. Most data are located on planetary surfaces, atmospheres, or interiors, and are described using specific coordinate frames.
- Data are also extremely diverse in nature: 1D/2D/3D/4D, bitmap vs vectorial, local vs event-related, observational vs laboratory vs modeling, celestial vs HR orbital images, remote sensing vs in-situ (or electro-magnetic vs particles)... Some of those call for particular display modes, in particular observations in reflected light or transmission, and particle measurements, have few counterparts in Astronomy.
- Variability: many data experience temporal variations at different time scales (secular, seasonal, local time). Therefore temporal coordinate systems must also be supported.

## DATA ACCESS PROTOCOLS / DATA MODEL

The user will write queries describing his search. Such queries must be translated in a standard form and used to search a catalogue of available data. This procedure is called a Data Access Protocol in the IVOA framework. Several protocols of interest for Planetary Science were already defined or under definition, addressing different contexts:

- IVOA protocols, which allow the user to search the data according to various criteria (i.e., Cone Search to locate objects near a position in the sky, TAP for tabular data…).
- The PDAP protocol, currently being defined in the IPDA framework. A typical use of PDAP is to address the entire contents of the PSA at ESA by querying their database globally, as a single data service. It therefore aims at describing the contents of the data, not only the observing conditions.
- The Spase-QL protocol (Space Physics Archive Search and Extract) defined by HDMC (Heliophysics Data and Model Consortium) used in plasma physics, e.g., to access the AMDA service at CDPP.
- VAMDC has also defined protocols focused on spectroscopy databases, in particular LineList.

These solutions were studied to figure out whether they could be extended to meet the needs of the Planetary Science VO. Most IVOA protocols appear to be related to objects located on the celestial sphere, but no solution currently exists to address data located e.g. on a planetary surface or in planetary atmospheres. Besides, the Cone Search protocol is marginally helpful for celestial images because it assumes the targets to have fixed coordinates on the sky, whereas Solar System objects are of course moving.

Conversely, PDAP potentially has the capacity to address most data of interest, provided topical extensions and perhaps some adjustments. However, PDAP is closely related to an implicit Data Model associated to the PDS3 format, therefore adapted to observational space-borne data only and providing little support for telescopic observations or laboratory data [3]. Since the main scope of PDAP is to access the space data archives, it was practically difficult to enlarge it to other purposes in a reasonable time frame. Another drawback is that IVOA data visualization tools make use of IVOA protocols, but not of the PDAP protocol. Most other protocols, including SPASE-QL, were not found versatile enough to support searches encompassing the full variety of Planetary Science data.

The outcome of this study therefore identified the need to develop a simple protocol based on existing ones, but including specific parameters permitting to handle Planetary Science data in general. The adopted solution was to define a restriction of the TAP protocol, called EPN-TAP. The TAP mechanism is here associated to a simple, specific Data Model (EPNcore); this is similar to the way the ObsTAP extension to TAP is associated to the Obscore data model. The Data Model is in practice a set of mandatory parameters describing a dataset, which is implemented in all data services and used as query parameters by the protocol [see EPN-TAP documentation, and companion paper]. Most of these parameters are related to the description of the data axes (time, coordinates, spectral), target, measurement type, and origin of the data. The protocol also states that quantitative parameters have to be provided in standard units in the catalogue, and that non-quantitative parameters (such as instrument names) are associated with identified reference lists. Most of these reference lists originate from the IAU or other independent sources such as NSSDC. This standardization makes it possible to send uniform queries to all EPN-TAP services, with no need for conversion or translation. Sets of optional parameters can be used to define topical extensions to the protocol, related to specific fields only (e.g. spectroscopy of minerals). Optional parameters are listed in the protocol and reserved for this use.

A more general Data Model has also been defined to describe the data of interest in a uniform way (Planetary Science Resource Data Model, or PSR-DM). This Data Model could be used in the future with a more general protocol. Such applications should use common metadata defined in the data model and the associated dictionary.

Specific domains may not be included in a single data model though, and must be handled with different systems (Fig. 2). For instance, other data models are being defined in relation with laboratory databases of solid spectroscopy (SSDM for the GhoSST data service) and atomic and molecular spectroscopy of gases (XSAMS), both in collaboration with the VAMDC consortium, which provides access to Atomic and Molecular Data. Access to services in these fields is better handled though their native protocols.

As in Astronomy, a potential difficulty to write valid queries is related to the many names of potential targets/objects. This problem is much worse for Solar System objects, which cannot be identified by a fixed position in right ascension / declination, and objects are often referred to via a number of names which are not always obvious (the official name of Halley's comet is for instance 1P, and asteroids are usually known under many names). A name resolver is therefore required to process user queries correctly. Such a name resolver has been devised at VO-Paris as part of the SSODnet project, and is available as a web service. It returns the official name of an object given any of its synonyms, with correct case.

## CATALOGUE / REGISTRY

Queries are sent to a global catalogue containing a description of all accessible data services and their capabilities. This description allows the system to make a first order selection of services matching the user's query. In the IVOA, this is done using a system of mirrored registries where data providers can record their services. Declaring a new data service in the registry system is the normal way to make it available in the VO, and to publicize it.

The registry is the core of the Planetary Science VO, as it makes it possible to discover all EPN-TAP compliant services in real time. It needs to be maintained on long time scales, and therefore the solution adopted was to use an IVOA compliant system. In particular it is based on the OAI-PMH standard and includes the use of the so-called "Dublin core" metadata and registry harvesting to allow multiple instances of registry information.

In addition, the solution adopted is to take advantage of TAPRegistryExtension to describe the EPN-TAP data model and data base schema information in the registry's VOServices declaration. VO-Paris has developed a Simple Rest Interface to the IVOA registry to provide easy access to all EPN-TAP services declared in the various IVOA registries.

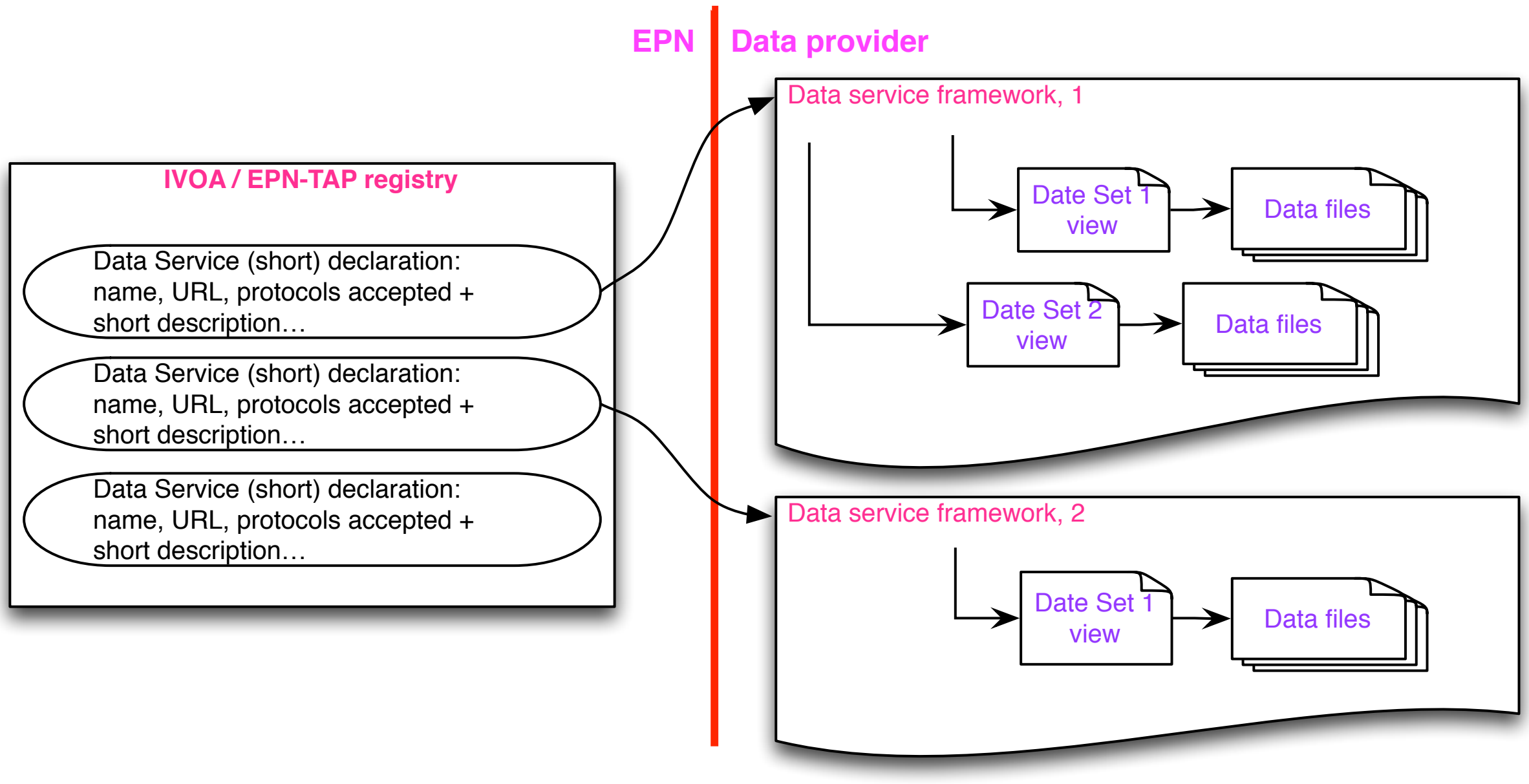

Figure 3: Europlanet registry and data services

We call EPN Service Data Model the data model used to describe the content of the Europlanet registry itself, i.e. all data services accessible through the Europlanet registry. The solution adopted for the Planetary Science VO follows the IVOA scheme, which is easy to maintain. Declarations of new services in the registry also follow the standard IVOA validation procedure.

The registry only contains a short description of the available data services, including their address, and the protocols they support. The detailed description of the services is maintained in the registry, but the data service itself is stored and maintained locally by the provider. The queries are therefore processed at two successive levels (Fig.3): services of interest are identified at the registry level, the query is passed to them according to the protocol indicated, and is processed locally. In this context, a data service is a series of data sets located in a given place, accessible through the same protocol with content described in a unique, local, catalogue.

Although Europlanet uses the IVOA registry standard to discover and select resources, other systems are used in neighboring fields. PDS4 defined a different type of registry to declare space archive products, which was later adopted by IPDA for PDAP services. SPASE and VAMDC registries follow yet other standards. Connecting these different registry systems will clearly be an important step forward in the future, e.g., mapping PDS4/plasma node and SPASE registries, or IPDA and Europlanet/IVOA ones, will enable access to more services in these environments.

# DATA ACCESS

The answer to a query is sent back by the service to the user. It typically provides a description, a link to the data, and information about data interface, but is in general limited to metadata and does not includes the data themselves. To be directly usable, this answer must also be formatted according to a known standard. This formatting can either be related to a data model (this is one of the solutions currently adopted by the VAMDC consortium) or remain at a general level (e.g. included in a VOTable). The EPN-TAP protocol states that the answer is returned in a VOTable, like PDAP and most IVOA protocols. In this case, a URL to the selected data product is usually included in the VOTable. In some cases when the data are directly integrated in the database, they can be embedded in the result VOTable and have to be parsed.

Reading the data themselves can be practically difficult, given the wide variety of formats used in Planetary Science. A special problem arises with PDS3 data (used by all current space data archives), for which no versatile reader is available. Although PDS4 is expected to greatly simplify this issue in the future [4], availability of "historical" space data in PDS4 is an open question. PDS3 data of many types can be read on-line, and then sent to IVOA tools using a mechanism set up at VO-Paris (LecturePDS library ran in an IDL/GDL shell, then transmitted through a JavaScript interface, see Demonstrators in Annex). IVOA tools handle some current data formats such as FITS or CSV, and TOPCAT now supports CDF that is commonly used by Plasma/Radio experiments to store temporal tables. Other formats are used in the Planetary Science community (such as KML, json, netCDF or HDF5), and their support is still under study.

# TOOLS / VISUALIZATION

Many visualization tools for basic data types were developed in the IVOA framework. The baseline for Planetary Science is to use existing tools, which are maintained by external teams; possible adaptations are then reduced to interface layers and support of various data formats, which may be discussed with the original software developers. The most flexible tools in this context appear to be Aladin and SAOImage/DS9 (for images and cubes), TOPCAT (for tables), SPLAT, Cassis, VOPlot, SpecView, and VOSpec (for vectors and spectra). These IVOA visualization tools all implement the SAMP data exchange protocol (based on XML-RPC, or preferably HTTP). Once the data are loaded in one of those tools, they readily become available to other tools through this protocol.

Specialized data types however may require specific visualization and measurement tools, which may be part of the data service user's interface. For instance, spectral laboratory data can be handled with precision in the GhoSST environment provided by IPAG together with its database [5]. Similarly, AMDA includes a specialized environment for plasma and solar Physics [6].

Many planetary data need to be projected in a particular coordinate system. Different situations may occur in Planetary Science:

- Sky coordinates (used e.g. for telescopic images of a target on background sky). This situation is routinely handled by Aladin and similar tools. The IMCCE Skybot service is accessible directly from Aladin to identify moving targets in a Field of View at a given time, e.g. to analyze telescopic plates.
- Planetary coordinates (e.g. for orbital images of planetary surfaces). This is similar to geographical coordinates on Earth, although the coordinate systems are body-related. Apart from the geometric computation (which is expected to lie on the data provider's side), plotting in such frames may also be an issue. High resolution imaging in particular requires a detailed description of planetary coordinates frames, including control point networks, in the Data Model, which is currently not available. Converters between coordinate systems may also need to be developed in the long run.
- Co-rotating coordinate systems are commonly used in some fields, e.g. for magnetospheres, and call for other types of conversions (one axis is then related to an external reference, typically the Sun's direction).

Among the functions of interest for Planetary Science, accurate registration of high resolution imaging data has a special importance. The specific coordinate systems need to be supported by image plotting tools, and we are therefore studying the possibly to add them in Aladin so that it can plot planetary maps produced by ISIS, for instance. This may require a slightly extension of the FITS format [7], e.g., the addition of a DATUM keyword. The IVOA STC (Space/Time Coordinates standard) includes some Solar System coordinate frames but is far for comprehensive, and it is set up in such a way that it is difficult to enlarge — for instance, it is currently impossible to describe the control point networks used on a planetary surface, the IAU planetary coordinate frames are not supported, and shape models of small bodies do not fit. The definition of an independent Solar System STC, based on IAU systems, therefore seems required. The WMS codes proposed by the IAU to describe coordinate systems [8], as well as the PDS unified planetary coordinates database [9] may answer this question.

A large fraction of our community works with Geographic Information Systems (GIS) to handle orthorectified data, either bitmap or vectorial. The GIS community has developed tools that use standards elaborated in the framework of the Open Geospatial Consortium (OGC), which can be used for Planetary Science [10] [11] [12]. A first demonstrator based on the Smart-1/AMIE images of the Moon has been developed at VO-Paris using the opensource QuantumGIS software, and the GIS-VO interface will certainly need to be developed in the near future.

Specialized tools for 3D plotting of Solar System bodies and spacecraft trajectories are also available, e.g. 3DView (developed by CNES, used by the IMPEx program [13]) or MATISSE (developed by ASI [14]). They both make use of JPL's Spice kernels to handle Solar System body shapes and motions. They are particularly adapted to representation of in-situ measurements, e. g. along a spacecraft trajectory.

## LINK TO COMPUTING ENVIRONMENTS

Services on data may also include standard computation/inversion algorithms. The question arises of the environment used to perform such operations, and their interface to the Planetary Science VO. Several possibilities already exist:

- The Aladin Java plug-in system allows developers to implement basic operations on images (such as computing the average spectrum in a region of interest).
- IDL can exchange data with Aladin in some environments via SAMP, and VOTables are supported to a certain extent via user libraries. Conversely, the only "complete" support for PDS3 data is within IDL/GDL. Compatibility with IDL/GDL would give access to a very wide range of data analysis applications in public IDL libraries. This would also provide a link with ENVI, which is widely used for surface studies and imaging spectroscopy.
- Some powerful libraries of image processing algorithms are available, such as the Orfeo Tool Box from CNES (developed for its Earth observation program). The French project Vahiné aims at providing a graphical interface to this library for remote sensing in Planetary Science [15].
- The IVOA has developed several protocols to handle computing on demand and automated workflows, e.g. UWS.

The implementation of workflows is scheduled in the near future. This will make it possible to develop processing pipelines running on calibrated data, with possible applications to on-going space experiments (data calibration, derived data…). In addition, web-based routines in javascript or python may be implemented to perform generic data processing functions on demand, relevant to image analysis.

## VESPA: THE EUROPLANET CLIENT

Queries to data services can be sent from a web form, from specialized software, or directly from visualization tools, as it is currently done in the IVOA system (e. g., from Aladin or TOPCAT). However, this requires the protocols to be implemented in the visualization tools, and therefore involves heavy developments.

The current solution for the Planetary Science VO is to use a web-based EPN-TAP client chaining all the functions described above [16]. The client is called VESPA, which stands for Virtual European Solar and Planetary Access [16]. It includes a user interface, queries/answers handling, and transmission to standard visualization tools through SAMP. Figures 4 and 5 illustrate these functions. The client first provides the user with an interface to write queries that are transmitted using the ADQL language. The IVOA registry is called to identify services containing data of interest, with a description of their interface. For each service, the answer is a VOTable providing a list of selected data files, which can be used either to restrain the query further, or to select data for quick-look. In this case, the data are transmitted to visualization tools via the common SAMP hub. The data files can also be passed to external, specialized environments for further processing. The SSODnet name resolver is used as a name completion service in the VESPA interface, both for Solar System objects and exoplanets. Thanks to the VOSI mechanism, the client may query all the parameters describing a TAP resource, beyond the mandatory parameters of EPNCore. VESPA also currently translates EPN-TAP queries and passes them to preselected PDAP services (ESA's PSA and JAXA's DARTS) to access PDS3 data from space agencies.

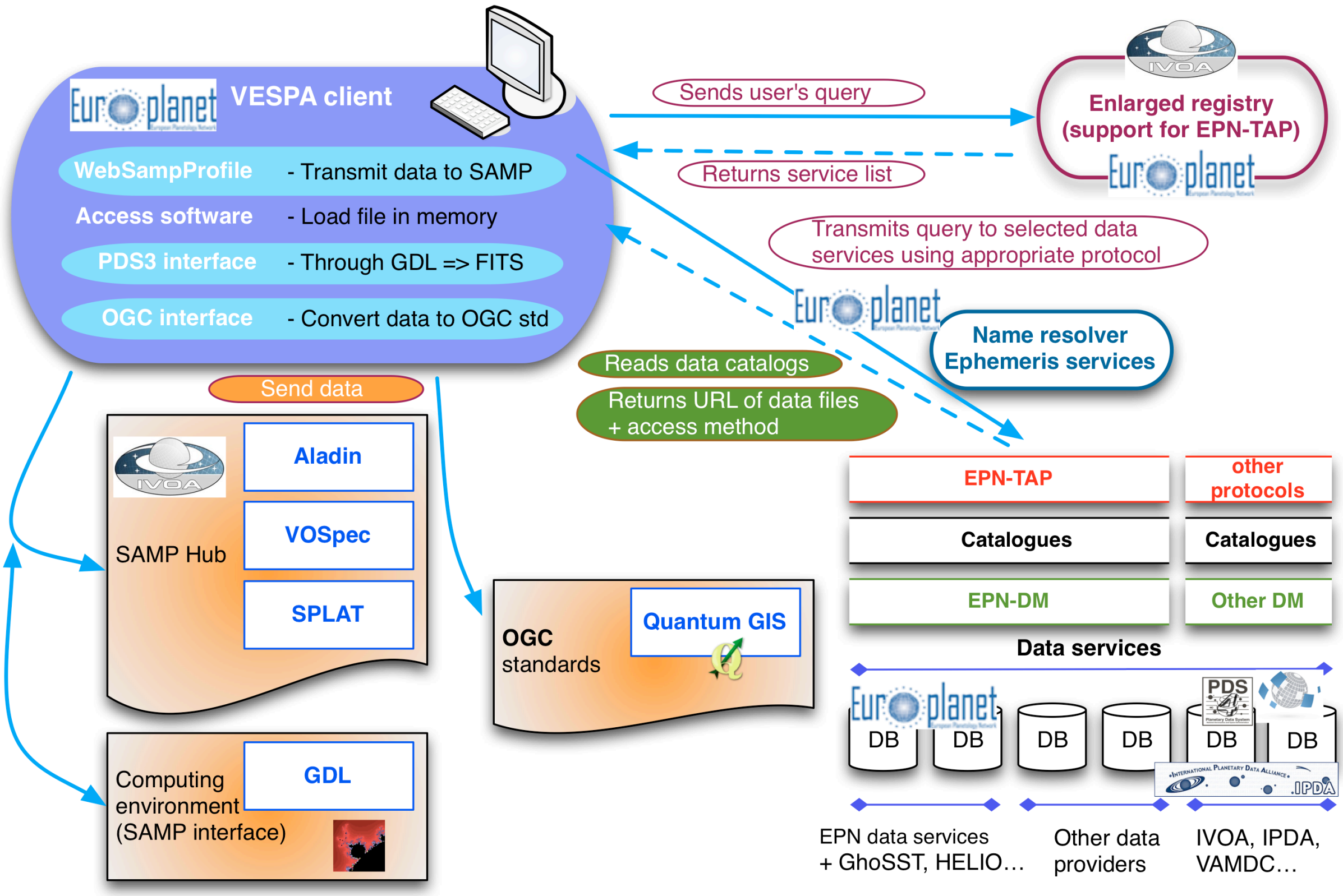

Figure 4: Functional diagram of the Planetary Science VO system, with indication of standards origin.

A difficulty encountered during the development of the system is the need to use proper spelling for several quantities, in particular target names (usually objects of the Solar System), which must follow IAU conventions. For this reason, the use of case-sensitive queries was mandatory, which is not the current standard of ADQL. The *ivo_nocasematch* function has been tested successfully when addressing the DaCHS framework on the data service side, but an update of ADQL will probably be required in the long term.

VESPA currently accesses the Planetary Science services declared in the IVOA registries, plus the PDAP services at ESA/PSA and JAXA/DARTS (those are currently test services, and do not necessarily implement the entire PDAP protocol). VESPA is currently the only EPN-TAP client available, and is therefore the main entry point to the Planetary Science VO (it also appears to be the only public PDAP client as of today). It is sufficient for users to search and retrieve data in the available databases. However, other clients can implement the EPN-TAP protocol to access the services described below. Data analysis services in neighboring fields can also implement EPN-TAP. This is being done in AMDA, which will get the ability to send automated queries to relevant EPN-TAP services documenting planetary auroras and radio measurements. This is also the case for the SILFE (Spectral Information of Low Frequency Emissions) radio data visualization tool prototype developed at LATMOS and LESIA.

Foreseen upgrades of VESPA in the near future include:

- The use of the name resolver to query services using all possible synonyms of a target.
- The ability to get composite results from several services and pass them to the appropriate tools.
- On-the-fly data conversion from PDS3 to FITS whenever required (currently a demonstrator).
- Implementation of a comprehensive VO-GIS interface (currently a demonstrator).
- The implementation of the VAMDC protocols to access gas spectroscopy services. Solid spectroscopy will be handled directly using EPN-TAP extensions.
- Development of a stand-alone EPN-TAP access package for implementation in external clients.

# DATA SERVICES

A dozen of EPN-TAP data services are already publicly available, and are listed in the Appendix below. Other projects are being finalized or studied.

New resources can be shared through this protocol by any data provider. The process (described in details in [17]) is the following:

- Set up an SQL database + install a framework supporting the TAP protocol (DaCHS and VO-Dance have been used successfully). A tutorial to install such a system is available on the VO-Paris Europlanet site.
- Ingest your data and create a view named epn_core describing all the "granules" (~individual data files). Pre-existing databases do not need to be changed, but the epn_core view must contain all the EPNCore mandatory parameters as expected: values of the descriptive parameters must be chosen from the reference lists, and quantitative data must be provided in the protocol's standard units.
- The data themselves can be either linked through the "access_url" parameter, or included in the epn_core view. Data formats are preferably FITS, VOTable, CDF, or ascii/csv for convenient handling by plotting tools.
- Set up the framework to provide answers as a VOTable with compliant content.
- Fill up an XML file describing the data service and submit it to an IVOA registry. Your service will be available from any existing Europlanet client, in particular from VESPA. It will also be accessible from TOPCAT directly (see doc on the VO-Paris Europlanet site).

Setting up a database is actually simple, since what is required may be limited to an SQL view in the DaCHS framework, and a local file archive. This system can even be used to provide evolved search functions in otherwise poorly organized archives: for instance, the INDEX file of a PDS3 archive can easily be converted in an epn_core view and link to data distributed on a space agency ftp server. This solution is being considered to provide powerful search capacities to the VIRTIS / Venus-Express archive in the PSA (imaging spectroscopy of Venus). Similarly, it can be used to link databases distributed on a simple web page, such as collections of asteroid spectra. Another possible application is to identify Planetary Science datasets inside large archives of telescopic observations, e. g. at ESO, where they are notoriously difficult to access. In such situations, the only step required is to build a consistent epn_core view from the database catalogue. The scheme described in Fig. 3 still applies, but the data/files provider and service provider are now completely independent.

Other services of interest for Planetary Science belong to the realm of IVOA but seem amazingly difficult to find. For instance, one would gladly use a simple service providing a list of the brightest stellar sources in various photometric bands, or reference stars, with corresponding spectra. Although such information is actually available in several observatories, it is currently difficult to access for the casual user.

A single DaCHS installation can accommodate several services; therefore grouping services is a time saving solution for small teams. DaCHs allows building EPN-TAP services from pre-existing databases, and prototypes of services have been built from external PostgreSQL and MySQL databases. Support can be provided at all these steps, and efficient tools are available to write the XML descriptors (e. g., the CDPP demonstrator in Appendix), to check service validity, and to register new data services.

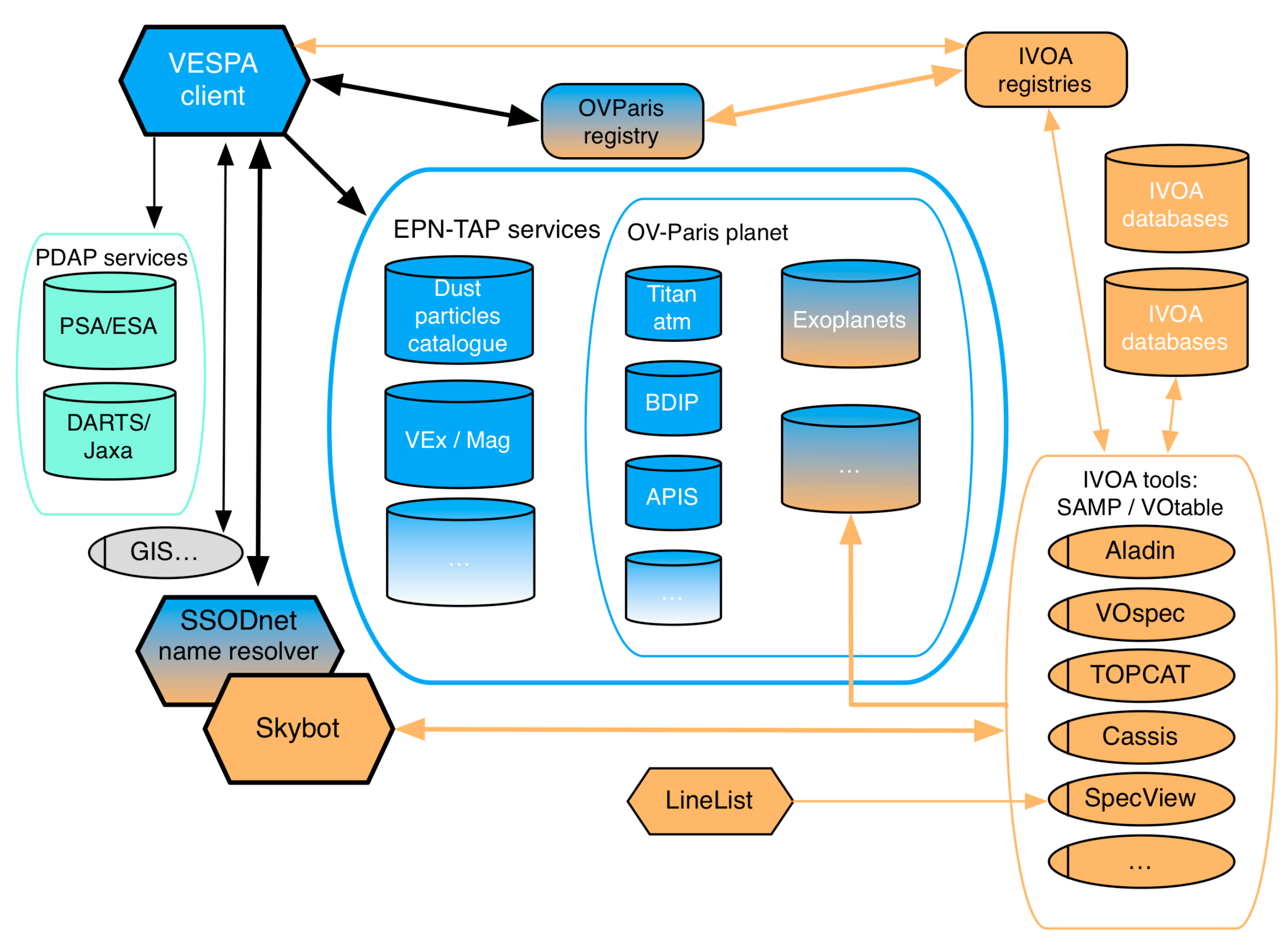

Figure 5: Service-centered view of the Planetary Science VO as of today. Europlanet elements are displayed in blue, IVOA's in orange, IPDA's in green, and OGC's in gray; multi-standard services are displayed in two colors. Only some examples of data services are displayed here.

The efficiency of the system for science purposes depends on the number of service connections, and therefore increases exponentially with the number of services connected in the same field. A typical use case is to compare Titan atmospheric profiles as retrieved from Cassini CIRS observations (Titan EPN-TAP service from LESIA) and simulated profiles from a Global Circulation Model (available at LMD, currently with no VO interface); areas where the model does not fit the observations could then be checked for regional radar or near-IR maps during the period, to look for evaporating lakes.

## PROSPECTS

Thanks to the Europlanet-RI program, the basic standards of the Planetary Science VO are defined and practical implementation has started. A registry system and a client have been developed to explore the data archives, and external VO tools are routinely used to provide on-line data visualization. Several test services are accessible and more are expected. A light framework and a procedure have been identified to allow small research teams to install data services, and two hands-on sessions have been organized since 2012 in Vienna and Paris, as well as tutorials during conferences (EPSC 2013 in London and EGU 2014 in Vienna). The next data services will focus on support to the on-going space missions, in particular Rosetta, Cassini, and future Mars missions.

Some technical issues are still being worked out however, for instance to improve the interface with the client on the one hand, the SSODnet name resolver and ephemeris systems on the other hand. The use of UCDs and Utypes associated with measurement parameters is also still in discussion. More integrated PDS3 support is also being developed, as well as adaptations of existing VO tools to planetary data (e. g., plotting reflectance spectra may currently be difficult). A list of standards used in the Planetary VO is provided in Figure 6.

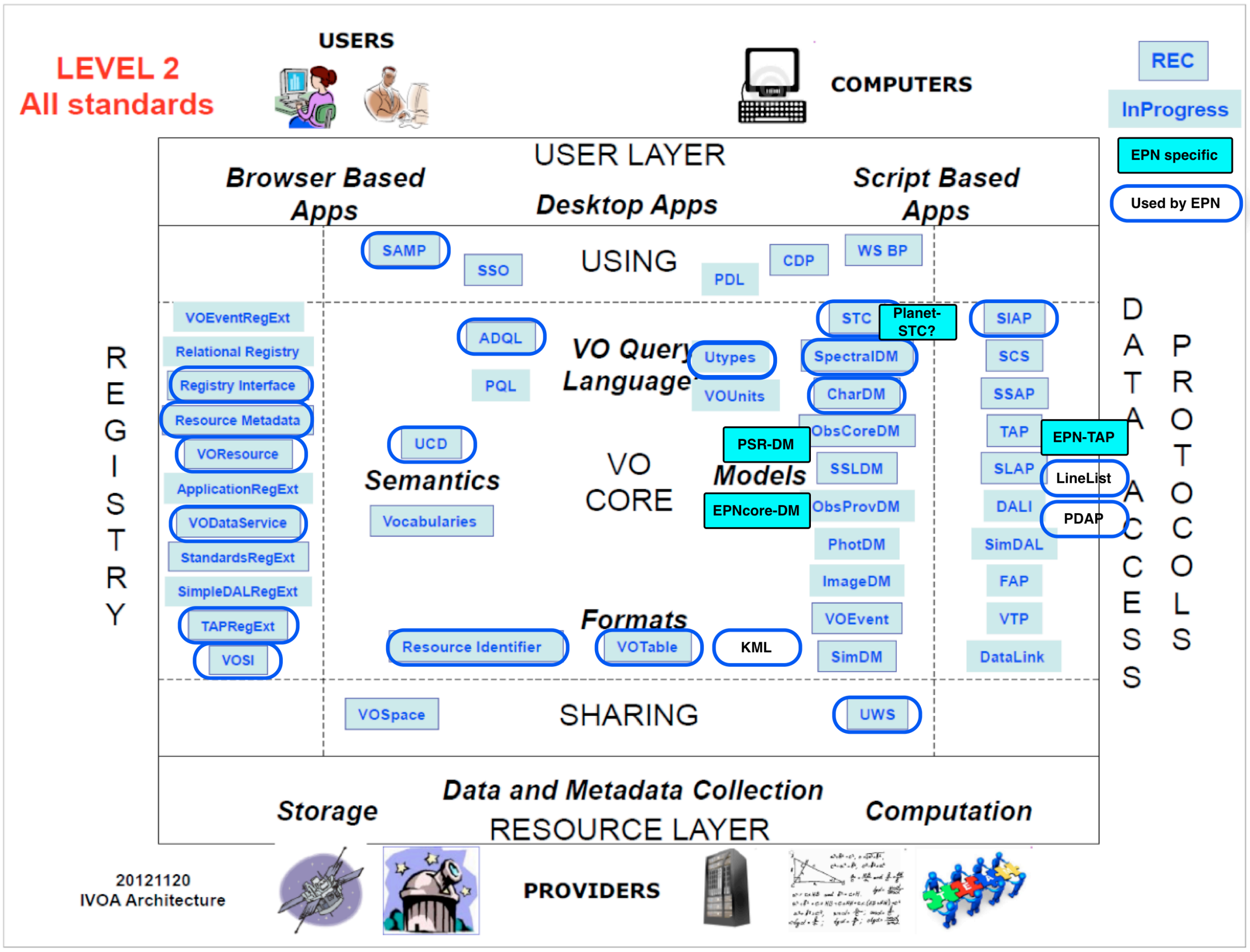

Figure 6: Standards used for Planetary Science, identified in the IVOA chart.
Europlanet-specific standards are displayed in blue, external ones in white.

Several more fundamental points are also still under study to facilitate data mining, in particular concerning the referencing of coordinate systems in use for the Solar System, and the identification of data sources (e.g., radio-telescopes and some orbital telescopes are not referenced by the IAU). Such points should be handled by a high-level authority such as IAU, or alternatively by consortia such as the IVOA or the IPDA. Planetary Science is more and more represented in IVOA activities, but the form this implication should take in the future is not yet very clear – either a liaison group with another entity (the IPDA or a « post-Europlanet » consortium), an « interest group » for Planetary Science, or direct participation of planetary scientists in many IVOA thematic working groups. This topic is being discussed in an IVOA/IPDA working group.

A VO-oriented sequel to the Europlanet-RI program is currently being defined in the Horizon 2020 framework. The participants would focus on providing additional data contents and setting public outreach activities taking advantage of the VO system. New participants will be welcome to contribute.

Data providers, including small research teams, are also invited to consider sharing their data in this system – the more data are available, the more attractive the Planetary Science VO is expected to become to science users. Another goal for this program will be to provide access to existing planetary data hidden in astronomy repositories, such as the ESO archive.

## Acknowledgements

The EuroPlaNet-RI project was funded by the European Commission under the 7th Framework Program, grant 228319 "Capacities Specific Programme". Additional funding was provided in France by the Association Spécifique Observatoire Virtuel / INSU.

# ANNEX

## Demonstrators

Demonstrators of the above mechanisms are available from the VO-Paris IDIS node:

http://voparis-europlanet.obspm.fr

- VESPA is currently the main entry point to the system. It sends queries to all EPN-TAP services declared in the VO-Paris registry, and to PDAP services.

  http://vespa.obspm.fr

  The “Custom resource” tab makes it possible to query a resource not yet declared, by providing the URL to its VO schema (this also allows querying any TAP service). The “Advanced query form” link in the service description of the result page allows the user to query all parameters from this service.

- Access to PDS3 data has been demonstrated using VIRTIS/VEx spectral cubes. A cube is read in an IDL (or GDL) session on the server and stored in FITS, the reference is sent to Aladin and plotted in X/Y coordinates. In Aladin, spectra are extracted on mouse click and forwarded to VOSpec or (preferably) SPLAT for plotting.

  http://voplus.obspm.fr/samp/SAMPWebProfile+FITS/demo.php

- A service to test and validate VOTables and data services is also proposed at VO-Paris. It can provide support to external contributors of the Planetary Science VO who wish to share data services.

  http://voparis-validator.obspm.fr/

- The SSODnet name resolver can be used to search a Solar System body by its different names, or by its coordinates at a given moment. Standard values are based on the IAU Minor Planet Center list. In VESPA, it is currently used as a completion service.

  http://vo.imcce.fr/webservices/ssodnet/?resolver

  The API is described here: http://api.voparis-tmp.obspm.fr/ssodnet/

Other demonstrators are available at CDDP:

- Access to planetary plasma data is implemented in the AMDA tool developed by the CDPP. Queries to distant databases (VEx/MAG in collaboration with IWF Graz, Cassini-MAPSKP…) are currently done using a SPASE-based connector:

  http://amda.cdpp.eu

- Access to external databases in AMDA has also been implemented using webservices (e.g., CDAweb). Current studies concern possible connections of PDS webservices with AMDA.

- A service to support writing of XML data file descriptors has been developed by CDPP. An interactive mode produces the XML files from a user-friendly web interface (which can be bypassed for pipe-line processing). This will soon be open for public use.

## First EPN-TAP data services

The first EPN-TAP services were designed as use cases to test the client/registry/visualization system in a variety of situations, and include:

• The Encyclopaedia of Extrasolar Planets (J. Schneider / VOPDC): reference compilation of published data maintained from 1995.

• APIS [Auroral Planetary Imaging and Spectroscopy] (L. Lamy, F. Henry / VOPDC): selection of HST UV observations, calibrated and projected, documenting auroral phenomena on giant planets.

• BDIP [Base de Données d'Images Planétaires] (P. Drossart, F. Henry / VOPDC): patrimonial archive of argentic telescopic images (from an older IAU project).

• IR spectroscopy of comet Halley (S. Erard, LESIA, IKI, PDS / VOPDC): calibrated dataset from the spectral channel of the IKS instrument on-board the Soviet Vega-1 spacecraft, 1986.

• Cosmic dust catalogue (IA2 & IAPS/INAF): a compilation of data from NASA particle catalogues 15 & 18.

• Basecom (J. Crovisier, M. Hirtzig / VOPDC): complete archive of cometary observations from Nançay radio telescope in the 18 cm region, with inferred water content and production rates (1982-2009, 53 comets).

• Vertical Profiles in Titan Middle Atmosphere (S. Vinatier / VOPDC): temperature and abundance profiles (100-500 km) from inversion of CIRS observations on Cassini, to be completed in 2014.

• M4ast (M. Birlan, M. Popescu, J. Normand / VOPDC): local archive of telescopic near-IR spectroscopy of asteroids (ground-based). Also includes a spectral fit tool from mineral databases.

• HELIO feature catalogue of active regions (J. Aboudarham, X. Bonnin / VOPDC): a catalogue of solar active regions retrieved from automated data analysis.

• HELIO feature catalogue of radio bursts (J. Aboudarham, X. Bonnin / VOPDC): a catalogue of solar type 3 bursts from Heliophysics feature catalogue.

• Jupiter routine observations (B. Cecconi, A. Coffre, E Thétas / VOPDC): dynamic spectra of Jupiter from the Nançay decametre array.

• Venus-Express / MAG instrument dataset (F. Topf / OEAW).

• A very basic service providing the masses and diameters of the planets, mainly intended for tutorials and demos (VOPDC). This may be enlarged to dwarf planets and satellites, and to other properties in the future.

## Acronyms & URL

ADQL — Astronomical Data Query Language http://www.ivoa.net/documents/latest/ADQL.html

AMDA — Automated Multi-Dataset Analysis at CDPP/Toulouse http://amda.cdpp.eu

CASSIS — Coordination Action for the Integration for Solar System Infrastructure and Sciences, a EU-funded program in FP7: http://cassis-vo.eu/

DaCHS — German VO (GAVO) Data Center Helper Suite http://docs.g-vo.org/DaCHS/

SCS — Simple Cone Search protocol from IVOA
http://www.ivoa.net/documents/latest/ConeSearch.html

EPNCore — Set of core parameters from EPN-DM, mandatory for EPN-TAP compatibility

epn_core — Name of a view in a database that provides the EPN-TAP parameters to be queried. Required for EPN-TAP compatibility.

EPN-DM — Specific Data Model to describe Planetary Science data in Europlanet-VO

| | |
|---|---|
| EPN-TAP | Specific protocol to access Planetary Science data in Europlanet-VO<br>http://voparis-europlanet.obspm.fr/xml/TAPCore/doc/html/index.html |
| Europlanet-RI | European Union funded program in FP7 (2009-2012) possibly extended in the 2016-2020 period<br>http://www.europlanet-ri.eu/ |
| FITS | (Flexible Image Transport System) one of the basic data formats in Astronomy, and a standard of the IVOA. |
| GhoSST | Grenoble Astrophysics and Planetology Solid Spectroscopy Thermodynamics. A solid-phase spectroscopy service including databases and tools, at IPAG/Grenoble. http://ghosst.osug.fr/ |
| GIS | Geographic Information Systems |
| HDMC | Heliophysics Data and Model Consortium |
| HELIO | HELiophysics Integrated Observatory, a EU-funded program in FP7: http://www.helio-vo.eu/ |
| IDIS | (Integrated and Distributed Information Services) Data handling activity in Europlanet-RI |
| IPDA | International Planetary Data Alliance https://planetarydata.org/ |
| IMPEx | Integrated Medium for Planetary Exploration, a EU-funded program in FP7:<br>http://impex-fp7.oeaw.ac.at |
| ISIS | (Integrated Software for Imagers and Spectrometers) A digital image processing software from USGS, commonly used to produce planetary maps from spacecraft observations. |
| IVOA | International Virtual Observatory Alliance http://www.ivoa.net/ |
| OGC | (Open Geospatial Consortium) Defines GIS standards |
| ObsCore | Set of core parameters from the Observation Data Model of IVOA http://www.ivoa.net/documents/ObsCore/ |
| ObsTAP | TAP protocol applied to the Observation Data Model of IVOA |
| PDAP | (Planetary Data Access Protocol) Protocol to access planetary data space archives, developed and maintained by IPDA (latest version dated 16/4/2013)<br>https://planetarydata.org/projects/previous-projects/copy_of_2011-2012-projects/PDAP%20Core%20Specification%20/pdap-v1-0-16-04-2013/ |
| PDS | Planetary Data System, the de facto standard for Planetary Science space borne archives<br>https://pds.jpl.nasa.gov/ |
| PDS4 | Coming PDS standard http://pds.nasa.gov/pds4/about/what.shtml |
| PSR-DM | Planetary Science Resource Data Model: a more exhaustive DM for Planetary Science, completing EPN-DM:<br>http://voparis-europlanet.obspm.fr/docs/PlanetaryScienceResource-DM-v2.0.pdf |
| SAMP | Simple Application Messaging Protocol from IVOA<br>http://www.ivoa.net/documents/SAMP/ |
| SPASE | Space Physics Archive Search and Extract http://www.spase-group.org/ |
| SSODnet | Solar System Open Database Network. A project at VO-Paris to connect selected databases with web page interface and provide them with an external VO interface. |
| TAP | (Table Access Protocol) One of the protocols developed by the IVOA to access astronomical data<br>http://www.ivoa.net/documents/TAP/ |
| UCD | (Unified Content Descriptor) Define measured physical quantities in the IVOA<br>http://www.ivoa.net/documents/latest/UCD.html |
| Utype | Description of data properties, in relation with a Data Model |

UWS Universal Worker Service pattern friom IVOA: http://www.ivoa.net/documents/UWS/

VAMDC Virtual Atomic and Molecular Data Center, a EU-funded program in FP7: http://www.vamdc.eu/

VESPA (Virtual European Solar and Planetary Access) Planetary Science VO project in the Europlanet2020 proposal. This is also the generic name of the EPN-TAP/PDAP client and main user interface: http://vespa.obspm.fr

VOSI Virtual Observatory Support Interface

http://www.ivoa.net/documents/VOSI/index.html